\documentstyle[11pt]{article}
\def\v1{\vspace{1cm}}
\def\be{\begin{equation}}
\def\ee{\end{equation}}
\def\bc{\begin{center}}
\def\ec{\end{center}}
\def\ik{\partial}
\def\vh{\varphi}

\newcommand{\bea}{\begin{eqnarray}}
\newcommand{\eea}{\end{eqnarray}}

\begin{document}

\title
{\bf Invariant Hamiltonian Quantization of General Relativity}
\author{
M. Pawlowski,\\
{\normalsize\it Soltan Institute for Nuclear Studies} \\
{\normalsize\it Warsaw,Poland.}\\[0.3cm]
V.N. Pervushin, V.I. Smirichinski \\
{\normalsize\it Joint Institute for Nuclear Research},\\
 {\normalsize\it 141980, Dubna, Russia.}\\
}

\date{\empty}

\maketitle

\medskip
PACS number(s):04.60.-m, 04.20.Cv, 98.80.Hw (Quantum Gravity)
\medskip

%\newpage

\begin{abstract}
{\small
{
The quantization of General Relativity invariant with respect to
time-reparametrizations is considered. We construct the
Faddeev-Popov generating functional
for the unitary perturbation theory in terms of invariants of the
kinemetric group of
diffeomorphisms of a frame of reference as a set of Einstein's observers
with the equivalent
Hamiltonian description ($t'=t'(t)$, $x'^i=x'^i(t,x^1,x^2,x^3)$).
The algebra of the kinemetric group has other dimensions than
the constraint algebra in
the conventional Dirac-Faddeev-Popov (DFP) approach to quantization.

To restore the reparametrization invariance broken in the DFP approach,
the invariant dynamic evolution parameter is introduced as
the zero Fourier harmonic of the space metric determinant.
The unconstrained version of the reparametrization invariant GR is obtained.
We research the infinite space-time limit of
the Faddeev-Popov generating functional  in the theory
and discuss physical consequences of the considered
quantization.
}}

\end{abstract}

\newpage

%\begin{multicols}{2}

\section{Introduction}

Quantization of General Relativity (GR) on the level of the unitary~\cite{f}
perturbation theory~\cite{d1}
was made by Faddeev and Popov~\cite{fp1,fp2} and by
DeWitt~\cite{dw} in accordance  with the
general Hamiltonian theory for constrained systems formulated by
Dirac \cite{d1,d2} and developed later by many authors (see e.g.
monographs \cite{ks,fs,ht}).
However, the construction of the unitary S-matrix in an infinite space-time
is not enough to answer the questions: What is Quantum
Gravity? and What is Quantum Universe with a finite measurable time
of its existence and a finite volume of its space?
Actual problems of the unification of elementary particle
physics with  General Relativity, cosmology of the Early Universe,
in particular, the description of quantum processes
at the beginning of the Universe
require the generalization
of these results ~\cite{fp2,dw} for a finite space-time.

In the present paper we try to generalize the Faddeev-Popov-DeWitt
construction of the unitary S-matrix for a finite space-time and to answer
the above mentioned questions about the quantum version of GR.

A main problem of generalization of this type is the invariance of GR with
respect to the group of general coordinate transformations.
This group includes the kinemetric group of diffeomorphisms of a
frame of reference as a set of Einstein's observers
with the equivalent
Hamiltonian description of GR \cite{vlad}.
Any Hamiltonian description  of GR (classical or quantum) should be invariant
with respect to transformations of this diffeomorphism group including
reparametrizations of the coordinate time.

Therefore, the coordinate time should be excluded from the
reparametrization invariant Hamiltonian dynamics.
Recall, that just the coordinate time is considered as the time of evolution
in both the Faddeeev-Popov-DeWitt unitary S-matrix and the Dirac Hamiltonian
approach to GR ~\cite{d1,d2,fp2}.

One of constructive ideas for restoration of the reparametrization-invariance
of the Hamiltonian dynamics in GR
is the introduction of the internal evolution parameter as one of
dynamic variables of the extendend phase space
\cite{Y,kuchar,torre,KPP,kpp,grg,plb,ps1}.
In the present paper, this idea is used to construct
the reparametrization-invariant
generalization of the Faddeeev-Popov-DeWitt unitary S-matrix in GR.
We fulfil the invariant Hamiltonian
quantizing GR with a dynamic evolution parameter identified with
the zero-Fourier harmonic of the space-metric
determinant ~\cite{grg,plb}.

This quantization is considered in the context
of the Dirac perturbation theory ~\cite{d1} in its simplest
version, without nontrivial topology, black holes, and surface terms
in the Einstein-Hilbert action.

The contents of the paper are the following.
In Section 2, we recall the conventional Dirac-Faddeev-Popov quantization
and present the invariant Hamiltonian approach to GR corresponding to the
diffeomorphism group of the Hamiltonian description.
Section 3 is devoted to the formulation of the Dirac perturbation theory
in GR with the invariant description
of classical dynamics, in the reduced phase space,
and of the measurable interval. In Section 4, we consider
the global sector of the invariant dynamics and define two different
standards of measurement of the invariant intervals:
the absolute standard for an Einstein observer and
the relative standard for a Weyl observer who can measure only
a ratio of the lengths of two intervals and who
treats GR as a scalar version of the Weyl conformal-invariant theory
\cite{plb,pr}.
Section 5 is devoted to the construction of quantum physical states
of the Dirac perturbation theory ~\cite{d1}.
In Section 6, we construct S-matrix  for
Quantum Universe and research the
conditions of validity of the conventional quantum field theory in
the infinite space-time limit.

\section{Invariant Hamiltonian dynamics of GR}

\subsection{The Dirac-Faddeev-Popov Quantization}

To state the problem considered in the present paper,
we recall the Dirac-Faddeev-Popov (DFP) quantized General Relativity
~\cite{d1,fp1,fp2,kp} which is given by the Einstein-Hilbert action
\be
\label{gr}
W^{gr}(g|\mu)= \int d^4x[-\sqrt{-g}\frac{\mu^2}{6} R(g) ] ~~~
(\mu^2=M^2_{Planck}\frac{3}{8\pi} )
\ee
and by a measurable interval
\be \label{dse1}
  (ds)^2_e=g_{\alpha\beta}dx^\alpha dx^\beta.
\ee
They are invariant with respect to general coordinate transformations
\be \label{x}
x_{\mu} \rightarrow  x_{\mu}'=x_{\mu}'(x_{0},x_{1},x_{2},x_{3}).
\ee
The Hamiltonian description of GR is fulfilled in the frame of
reference with
the Dirac-ADM $3+1$  parametrization of the metric components \cite{ADM}
\be
\label{dse}
  (ds)^2_e=g_{\mu\nu}dx^\mu dx^\nu= N^2 dt^2-{}^{(3)}g_{ij}\breve{dx}{}^i
  \breve{dx}{}^j\;~~~~~\;\;(\breve{dx}{}^i=dx^i+N^idt) ,
\ee
and it is given by
the first-order representation of the initial action ~(\ref{gr})
\be \label{Egr}
W^E=\int\limits_{t_1}^{t_2} dt \int d^3x
[-\pi_{ij}\dot q^{ij} - N_q {\cal H}
- N^i {\cal P}_i + \mbox{surf.terms.}]
\ee
obtained by Dirac ~\cite{d1} in terms of harmonical variables
\be \label{hv}
q^{ik}=||g|| g^{ik}, ~~~ N_q=N ||g||^{-1/3},~~~~~~
 (||g||=\det({}^{(3)}g_{ij}),~~~q=||q^{ij}||) ,
\ee
\be \label{hh}
{\cal H}=\frac{6}{\mu^2} q^{ij} q^{kl}[\pi_{ik}\pi_{jl}-\pi_{ij}\pi_{kl}]
+\frac{\mu^2 q^{1/2}}{6}{}^{(3)}R,
\ee
\be \label{ph}
{\cal P}_i=2[\nabla_k(q^{kl}\pi_{il})-\nabla_{i}(q^{kl}\pi_{kl})].
\ee
We omit in equation ~(\ref{Egr}) surface terms and use a finite space-time
\be \label{fst}
\int d^3x=V_0,~~~~~~~~~~~ t_1 < t < t_2.
\ee
According to the Dirac general Hamiltonian theory ~\cite{d1,d2},
the time components of the metric $N_q, N^i$ are considered as
the Lagrange multipliers,
and six space components $q^{ij}$, as dynamic
variables.
Four local Einstein equations for the time components of metric
\be \label{grconstr}
 \phi_0:={\cal H} = 0;~~~~~ \phi_i:={\cal P}_i = 0.
\ee
are treated as the first class constraints which remove four variables from
the extended phase space.
These four constraints should be supplemented by
four local  gauges which remove their canonical partners.
Dirac \cite{d1} imposed the "gauge"  constraints
\be \label{dc}
\chi_0:=\pi := q^{ij}\pi_{ij} = 0,~~~~~\chi^k:=\partial_l(q^{-1/3}q^{lk}) = 0
\ee
to express the initial action in terms of two
independent dynamic variables of the metric.

The algebra of commutation relations of all constraints forms
the Faddeev-Popov (FP) determinant~\cite{fp1,fp2} which restores the
unitarity of S-matrix in quantum theory.

This  determinant for constraints~(\ref{grconstr}),~(\ref{dc})
has been computed in the monograph~\cite{kp}
\be \label{det}
det\left\{ \phi_{\mu},\chi_{\nu}\right\} = det A det B,
\ee
where $A$ and $B^i_k$ are operators acting by the rules
\be \label{ca}
Af=q^{ij} \nabla_i  \nabla_j f + q^{1/2}{}^{(3)}R f;
\ee
\be \label{cb}
B^i_k \eta_i=q^{-1/3}q^{lj} [\delta^i_k
\partial_l \partial_j +
\frac{1}{3}\delta^i_l \partial_j \partial_k] \eta_i .
\ee
According to the Faddeev-Popov prescription~\cite{fp1},
the generating functional of Green functions has the form of
the functional integral
\be \label{dfp}
Z_{DFP}[t_1|t_2]=\int D(q,\pi,N_q,N^k) [FP]_s[FP]_t
\exp\left\{ iW^E(g|\mu)+\mbox{sources}\right\},
\ee
where
\be \label{prod}
D(q,\pi,N_q,N^k)=
\prod\limits_{x}\left(\prod\limits_{i<k} dq^{ik}d\pi_{ik}
\prod\limits_{j=1 }^{3 }dN^j dN_q \right)
\ee
and
\be \label{fps}
[FP]_s=\delta(\chi_{j})) det B.
\ee
\be \label{fpt}
[FP]_t=\delta(\chi_{0})) det A.
\ee
are the space and time parts of the FP determinant.
This functional and its gauge-equivalent versions are the foundation of
the quantum field theory approach to General Relativity
in the framework of the perturbation theory formulated in the
infinite space-time limit.

\subsection{Diffeomorphism group of the Hamiltonian description}

To extract any physical information from relativistic systems,
one should point out a frame of reference~\cite{vlad}. The latter
means to answer the questions: Which quantities can an observer measure?
How do these quantities connect with the metric components?
How do results of measurements depend on a state of motion of an
observer?
The simplest example is the description of the energy spectrum of
a relativistic particle in Special Relativity (SR) with the Poincare
group of symmetry.
There are two distinguished frames of the Hamiltonian description:
the rest frame of
an observer, and the comoving frame. In both the cases, to solve the
physical problem and to obtain
the spectrum, it is sufficient to restrict SR by only the subgroup of
the Poincare group which exludes the pure Lorentz transformations.
The latter are needed to answer the last question: How does the
spectrum in the rest frame connect with the one in the comoving frame?

The main assertion of the present paper is the following:
for the Hamiltonian description of GR in a definite frame of reference,
it is sufficient to restrict GR by the group of diffeomorphisms
of this frame.

Recall that
the Hamiltonian formulation ~(\ref{Egr})
is based on the possibility  to introduce a set of
the three-dimensional
space-like hyperspaces~(\ref{dse}) numerated by the time-like coordinate
in the four-dimensional manifold of the world events.
This set can be defined
within transformations of a kinemetric subgroup of
the group of general coordinate transformations (\ref{x})
\cite{vlad,grg,plb,ps1}
\be \label{gt}
t \rightarrow  t'=t'(t);~~~~~
x_{i} \rightarrow  x_{i}'=x_{i}'(t,x_{1},x_{2},x_{3}),
\ee
which includes one global function (the time reparametrizations $t'(t)$)
and three local ones ($x_{i}'(t,x)$). This is the group of diffeomorphisms
of a set of Einstein's observers with the equivalent Hamiltonian dynamics
~(\ref{Egr}). This continuum of "observers" with the diffeomorphism group
~(\ref{gt}) is called  the kinemetric frame of reference~\cite{vlad}.
Action ~(\ref{Egr}) can be written in terms of the diffeomorphism invariants
\be \label{Edi}
W^E(g|\mu)=\int\limits_{t_1}^{t_2} dt \int d^3x
[-\pi_{ij}D_t q^{ij} - N_q {\cal H}],
\ee
including invariant differentials
\be \label{difin}
dt D_t q^{ij}=dt [\dot q^{ij} + q^{ik} \nabla_k N^j + q^{jk} \nabla_k N^i
-2q^{ij}  \nabla_k N^k].
\ee
The dimensions of diffeomorphism group~(\ref{gt}) do not coincide
with the dimensions of the Dirac-Faddeev-Popov (DFP) algebra of constraints.

It is easy to see that,
in the case of a finite time interval $(t^1 < t < t^2)$,
the DFP generating functional ~(\ref{dfp})
breaks the invariance with respect to the global reparametrization of the
coordinate time ~(\ref{gt}).

On the other hand,
a local transformation of the coordinate time $t'=t'(t,x)$
 goes beyond the scope of this group ~(\ref{gt}).

Thus, the generating functional ~(\ref{dfp}) takes into account
the symmetry which is absent in the diffeomorphism group ~(\ref{gt})
and it breaks the symmetry contained in this group.

Below we show that three gauges~(\ref{dc}) $\chi^k=0$ are sufficient
to remove all local ambiguities from the Hamiltonian dynamics, so that
the fourth local gauge can contradict the Hamiltonian equations of motion.
Only the space integral from the equation for the lapse-function could
be considered as the standard first class constraint
(accompanied by the second class one, i.e. gauge)
in agreement with the diffeomorphism group of the
Hamiltonian description~(\ref{gt}).

\subsection{The invariant Hamiltonian scheme}

To restore the time-reparametrization invariance, we introduce
the internal evolution parameter
as the zero Fourier harmonic $\vh_0(t)$  of
the space metric determinant logarithm~\cite{grg}.
This evolution parameter can be extracted
by the conformal-type transformation of the metric
\be \label{conf}
g_{\alpha\beta}(t,x) = (\frac{\vh_0(t)}{\mu})^2\bar g_{\alpha\beta}(t,x) .
\ee
The local part of momentum of the space metric
determinant $\bar \pi$ and its motion equation equivalent $\bar k$
\be \label{cs1}
\bar \pi (t,x) := \bar q^{ij}\bar \pi_{ij},~~~~~
\bar k (t,x):=\frac{{\bar q_{ij}D_t \bar q^{ij}}}{\bar N_q}=
 \frac{{D_t\log \bar q}}{\bar N_q}
\ee
are given in the class of functions with
the non-zero Fourier harmonics, so that
\be \label{cstr2}
\int d^3x \bar \pi (t,x)=0,~~~~~~~~~~~
 \int d^3x \bar k (t,x) = 0.
\ee
Using the transformational properties of the curvature $R(g)$ with respect to
the transformation~(\ref{conf}) it is easy to obtain the action~(\ref{gr})
in the form
\be \label{Econf}
W^E(g|\mu)=W^E(\bar g|\vh_0) -\int\limits_{t_1 }^{t_2 }dt \vh_0
\frac{d}{dt}(\dot \vh_0 V)
~~~~~~~~~~~~(V=\int d^3x {\bar N_q}^{-1}),
\ee
with the same number of variables.
The Hamiltonian form of this action is
\be \label{Egrc}
W^E(q|\mu)=\int\limits_{t_1}^{t_2} dt \left(\int d^3x
[-\bar \pi_{ij}\dot {\bar q}^{ij} - \bar N_q \bar {\cal H}
- N^i \bar {\cal P}_i]
-\dot \vh_0P_0 +\frac{P_0^2}{4V} \right),
\ee
where the densities of the local excitations
$\bar {\cal P}_i$ and $\bar {\cal H}$
repeat the conventional Einstein ones~(\ref{ph}) and~(\ref{hh})
where the Planck constant $\mu$ is replaced by the internal evolution
parameter $\vh_0$ and  $q,\pi$ are
replaced by $ \bar q, \bar \pi$
\be \label{bar}
\bar {\cal P}={\cal P}(q,\pi \rightarrow \bar q, \bar \pi);~~~~
\bar {\cal H}={\cal H}(q,\pi \rightarrow \bar q,
\bar \pi;~\mu \rightarrow \vh_0).
\ee
We shall consider action~(\ref{Egrc}) as one of
the kinemetric invariant versions of the Hamiltonian  dynamics with
the global variables which
allow us to extract the time-reparamerization invariant physical
consequences in accordance with the diffeomorphism group~(\ref{gt}).

\subsection{Unconstrained form of GR}

The unconstrained form of GR is obtained by explicit resolving the constraints.
The space constraints
\be \label{echn}
\frac{\delta W^E}{\delta N^k}=0\,
\Rightarrow\,
\bar {\cal P}_k = 0
\ee
and the diffeomorphism group~(\ref{gt}) allow us
to remove from the extended phase space three local components
of the graviton field by
fixing the  gauge ~(\ref{dc})
\be \label{dgh}
\chi^k=\partial_i (\bar q^{-1/3}\bar q^{ik})=0.
\ee
Constraints ~(\ref{echn}), and ~(\ref{dgh})
can be explicitly solved by the decomposition of momenta $\bar \pi_{ij}$
into the transverse part $\bar \pi^T_{ij}$
and longitudinal components $f_k$
\be \label{pr}
\bar \pi_{ij}=
\bar \pi^T_{ij}+
 \bar q^{-1/3}[\partial_i f_j + \partial_j f_i
-\frac{2}{3} \bar q_{ij}\bar q^{lk} \partial_l f_k ]~~~~~~
(~\partial^i[\bar q^{1/3}\bar \pi^T_{ij}]=0~).
\ee
Substitution of this decomposition into constraint~(\ref{echn})
leads to the equation for the longitudinal component $f_k$
\be \label{bcr}
 \bar {\cal P}_i(\bar \pi, \bar q) =
 \bar {\cal P}_i(\bar \pi, \bar q) -\frac{1}{2}B^k_i f_k= 0 \Rightarrow
B^k_i f_k=
\frac{1}{2}\bar {\cal P}_i(\bar \pi^T, \bar q)
\ee
in the class of local functions with the non-zero Fourier
harmonics where the reverse operator $(B^{-1})_k^i$  ~(\ref{cb}) exists.

Let us consider the equation for the lapse function
\be \label{ech}
\bar N_q\frac{\delta W^E}{\delta \bar N_q}=0\,
\Longrightarrow\,
\frac{P_0^2}{4V^2\bar N_q}-\bar N_q\bar {\cal H}=0.
\ee
The integration of of equation~(\ref{ech}) over the
space coordinates determines the global momentum $P_0$ in action~(\ref{Egrc})
\be\label{glop}
\frac{1}{V}\left(\int d^3x\bar N_q\frac{\delta W^E}{\delta \bar N_q}\right) = 0\,
\Longrightarrow\,
(P_0)_{\pm}=\pm 2 \sqrt{V H}\equiv\pm H^R
~~~~~~(~H=\int d^3x'{\bar N_q}\bar{\cal H}~).
\ee
as the functional of all other variables.
It is the generator of evolution with respect to the
dynamic evolution parameter $\vh_0$.
Thus, the global part of equation~(\ref{ech}) removes
the global momentum $P_0$, in the correspondence with
the diffeomorphism group~(\ref{gt}).
The orthogonal to~(\ref{glop}) (local) part of the same equation~(\ref{ech})
\be\label{locp}
\left(\bar N_q\frac{\delta W^E}{\delta \bar N_q} -
\frac{1}{V}\int d^3x\bar N_q\frac{\delta W^E}{\delta \bar N_q}\right) = 0~
\Longrightarrow~
{\bar N_q} \bar {\cal H} - \frac{ H}{\bar N_qV}=0,
\ee
together with six equations for the transverse
space metric components $\bar \pi^T_{ij},~\bar q^{ij}$,
determines the lapse function $\bar N_q$ with
an arbitrary time-dependent factor $\beta(t)$
\be \label{lapse}
\bar N_q(t,x) = [\sqrt{\bar {\cal H}(t,x)}]^{-1} \beta(t).
\ee
This factor represents the Lagrange multiplier.
The generator of the dynamic evolution~(\ref{glop})
\be \label{gde}
H^R = 2 \sqrt{V H}=2 \int d^3x \sqrt{\bar {\cal H}(t,x)}
\ee
does not depend on this factor $\beta(t)$.

In accordance with the diffeomorphism group~(\ref{gt}), we  consider
as constraint only
equation~(\ref{glop}) which is the equation for the Lagrange multiplier
$\beta(t)$.

The global constraint~(\ref{glop}) has two solutions which
correspond to two reduced systems with the actions
\be
\label{d7}
W^R_{\pm}=\int\limits_{\vh_1=\vh_0(t_1)}^{\vh_2=\vh_0(t_2)}
d\vh \left\{\left(-\int
d^3x \bar \pi^T_{ij}\ik_{\vh}\bar q^{ij}\right)\mp H^R\right\}
\ee
where $H^R$ is the Hamiltonian of evolution~(\ref{gde}) of the
reduced phase space
variables $(\bar \pi^T_{ij},~\bar q^{ij})$ with respect to the
dynamic evolution parameter $\vh=\vh_0$.

Following to Dirac~\cite{d2}, we call the sector
of the reduced phase space described by action ~(\ref{d7})
the Dirac "observables".
These variables are kinemetric  invariants by the construction.
The equations of motion of the reduced unconstrained system are
\be \label{pi}
\frac{\delta W^R}{\delta \bar \pi_{ij^T}}=0 \Rightarrow
\frac{\partial  \bar \pi_{ij^T} }{\partial \vh}=
\pm\frac{\delta H^R}{\delta \bar q^{ij}} ,
\ee
\be \label{qi}
\frac{\delta W^R}{\delta  \bar q^{ij} }=0 \Rightarrow
\frac{\partial  \bar q^{ij} }{\partial \vh}=
\mp\frac{\delta H^R}{\delta \bar \pi^T_{ij}} .
\ee
Solutions of equations~(\ref{pi}),~(\ref{qi})
determine the dependence
of the Dirac observables with on the dynamic evolution
parameter $\vh$.

The main problem is to construct the time-reparametrization invariant
Faddeev- Popov generating functional for the unitary perturbation theory.

\section{The invariant version of the Dirac perturbation theory}

The reparametrization-invariant version of the perturbation theory
begins from the nonperturbative background metric with the homogeneous part
of the space metric~(\ref{conf}) (which gives the dynamic
evolution parameter) and the global component of the lapse
function $\bar N_0$~(\ref{lapse})
which defines the reparametrization-invariant conformal
time
\be \label{time}
\bar N_q=N_0 \bar N;~~~~~~~dT=N_0(t)dt ~~~~~~~~~~~~  (dT'=N_0'dt'=dT).
\ee
For the local part of metric, we  use the version of
the Dirac perturbation theory~\cite{d1}
\be \label{pt}
{\bar q}^{1/3}{\bar q}_{ij}=\delta_{ij}+  h^T_{ij};
~~~\bar q^{1/3}=1+4 z+...;~~~~~
 \bar N=1 + \nu +... .
\ee
Asymptotic states will be considered in the neglect of
interactions, in  accordance with  the standard suppositions of quantum
field theory.

In the lowest order of this theory
equation ~(\ref{locp}) determines $z = (\log \bar q)/12$
\be \label{nopt}
-\frac{2\vh^2}{3}\Delta z =
\bar {\cal H}_0- \rho_0~~~~~
(~\rho_0=\frac{\int d^3x \bar {\cal H}_0}{V_0};~~~~{V_0=\int d^3x}~)~.
\ee
A solution of this equation recalls the
FP gauge~\cite{fp2} where the internal evolution parameter is changed
by $\mu$ and instead of the massive matter we have the non-zero
Fourier harmonic part of
the Hamiltonian for two transverse
and trace-less gravitons
\be \label{pert}
\bar {\cal H}_0:=
\frac{6}{\vh^2}(\pi^T)_{ij}^2+ \frac{\vh^2}{24}(\partial_k (h^T)_{ij})^2
:=\vh^{-2} \bar {\cal H}_K+\vh^2\bar {\cal H}_R.
\ee
The solutions of equation~(\ref{nopt}) is usually  treated
as the Newton interaction of particles, i.e. of the transverse and trace-less
gravitons $h^T_{ij}$ which  form the asymptotical physical states, similar
to photons in QED.
These transverse gravitons $h^T_{ij}$ are considered in paper ~\cite{ps1}
in context of the invariant Hamiltonian quantization.
The trace component
of the graviton momentum $\bar q^{ij}\bar \pi_{ij} = p_z/12$ disappears
from the kinetic part of the Hamiltonian ~(\ref{pert}) $\bar {\cal H}_K$
as a result of the solution of the space constraints.
Nevertheless, the momentum $p_z$ is not equal to zero as it follows from the
equations of the initial extended system for $p_z$
\be \label{pzpt}
2\vh^2(z'-\partial_k \bar N_k)=p_z~~~~~~
   (z'=\frac{\partial z}{\partial T}=\sqrt{\rho_0}\partial_{\vh}z),
\ee
and for $\bar N^k=N^k/N_0$
\be \label{nkpt}
-\frac{2\vh^2}{3}(\partial_k z'-\Delta \bar N_k)={\cal P}^T_k:
=(\pi^T)_{ij} \partial_k (h^T)_{ij},
\ee
in  contrast with the Dirac gauge~\cite{d1}.
The perturbation part of the lapse function $\nu$ is determined from
the motion equation of the reduced system for $z$  unambiguously
\be \label{zpt}
\nu=\frac{2\vh^2}{3} \Delta z \rho_0^{-1}~.
\ee
One can see that the range of applicability of the Dirac perturbation
theory ~(\ref{pt}) is the region where derivatives
are far less than the internal evolution parameter $ \Delta f/f << \vh^2$.

In the opposite limit $\vh_0~\rightarrow~0$,
we got the local version of the model of an
anisotropic universe considered by Misner ~\cite{M}.

\section{Measurable quantities}

\subsection{Geometry}

The dynamic sector of the unconstrained GR restricted by the
Dirac "observables"~(\ref{d7}),~(\ref{pi}),~(\ref{qi})
is not sufficient to
determine evolution of the Einstein invariant interval
\be \label{eint}
(ds_e)^2=(\frac{\vh}{\mu})^2 {\bar q}^{1/6}(ds_c)^2,
\ee
here $ds_c$ is the conformal invariant interval
\be \label{wint}
(ds_c)^2={\bar q}^{1/3}[dT^2 {\bar N}^2 -
{\bar q}_{ij}\breve dx^i \breve dx^j];~~~(\breve dx^j=dx^j+\bar N^jdT)
\ee
which does not depend on the global variable $\vh$.
These intervals characterize the measurable geometry of the space-time and
contain the shift vector $\bar N^k$ and  invariant
time parameter~(\ref{time}).
The latter is well-known in the classical Friedmann cosmology ~\cite{KPP,plb}
as the conformal time connected with the world Friedmann time by
the relation
\be \label{world}
dT_f=\frac{\vh(T)}{\mu}dT.
\ee
Measurable geometrical quantities go out from the set of the Dirac
"observables",
and can be determined by  invariant
equations of the initial extended system for the global variables
$P_0, \vh$ and the local ones $\bar \pi_{jl}$
which are omitted by the reduced action~(\ref{d7}).
In particular, the evolution of the Universe is not also included
in the dynamic sector of the Dirac "observables".

\subsection{Evolution of a universe}

The evolution of a universe is the dependence of the measurable time
~(\ref{time}) on
the internal evolution parameter given by the equation of the extended
system
\be \label{nono}
\frac{\delta W^E}{\delta P_0}=0\,
\Rightarrow\,
\left(\frac{d\vh}{dT}\right)_{\pm}
=\frac{(P_0)_{\pm}}{2V}
=\pm\sqrt{\rho({\vh})};~~~\rho=\frac{\int d^3x \bar {\cal H}}{\bar V_0}=
\frac{\bar H}{V_0};
\ee
The integral form of the last equation
\be
\label{70}
T({\vh_0})=\int\limits_0^{\vh_0}d\vh
{\rho}^{-1/2}(\vh).
\ee
is well-known as the Friedmann-Hubble law
in the Friedmann-Robertson-Walker cosmology.
It is natural to call the Hamiltonian $\bar H$  the "measurable" one, as it
determines the evolution of the Dirac observables with respect to
the measurable time $T$
\be \label{mt}
   f':=\frac{\partial f}{\partial T}
=\sqrt{\rho}\partial_{\vh}f = \left\{\bar H, f\right\}.
\ee
Another global equation of the extended system
\be
\label{d55}
\frac{\delta W^E}{\delta {\vh}}=0
\,\Rightarrow\, P'_0=V\frac{d}{d{\vh}}{\rho}({\vh})
\ee
leads to the conservation law for the measurable Hamiltonian
$\bar H$~\cite{grg}
\be \label{conl}
\vh^{-2} \bar H'_K + \vh^2 \bar H'_R = 0,
\ee
where symbols $K,R$ mean the kinetic and potential parts
(see eq.~(\ref{pert})).
The shift vector is determined by the equation
\be \label{shift}
 \frac{\delta W^E}{\delta \bar \pi_{jl}}=0\,
\Rightarrow\,
\partial_T {\bar q}^{jl}+ \nabla^j \bar N^l+\nabla^l \bar N^j=
\frac{12 \bar N}{\vh^2}
(\bar q^{ij} \bar q^{kl}\bar \pi_{ik}-\bar q^{jl} \bar \pi).
\ee

\subsection{Standards of measurement}

As it was shown in papers ~\cite{grg,plb}, GR with the Einstein-Hibert action
can be also treated as the scalar version of the Weyl
conformal theory~\cite{pr} with the scalar field $\Phi_w$
considered as the measure of a change
of the length of a vector in its parallel transport. In this case the role
of the metric scale field $\vh_g=\mu q^{1/12}$ in GR is played by
the Lichnerowich~\cite{L,Y} conformal invariant variable
$\vh_c=\Phi_w q^{1/12}$  of the scalar field.
Dynamics of both the Einstein GR and the Weyl theory is the same
(including the matter sector where the scalar field forms  masses of fermion
and boson fields),
but not standards of measurement. An Einstein observer measures the
absolute lengths $(ds)_e$, while a Weyl observer  can measure only
the ratio of lengths of two vectors $(ds)_w=(ds_1)_e/(ds_2)_e=(ds_1)_c/(ds_2)_c$
which is conformal-invariant.
Thus, a classical state of the universe in GR
(with the Einstein-Hilbert action)
is determined both by the dynamic sector of the
Dirac "observables" in the reduced phase
space and  the geometrical sector of "measurables"; the latter are determined
not only by invariant dynamics of the Einstein-Hilbert action , but also by
standards of measurements.

The same GR dynamics
corresponds to different cosmological pictures for different
observers:
an Einstein observer, who supposes that he measures an absolute interval,
obtains the Friedmann-Robertson-Walker (FRW) cosmology where the
red shift is treated as expansion of the universe;
a Weyl observer, who supposes that he measures a relative interval $D_c$,
obtains the Hoyle-Narlikar cosmology \cite{N}.
The red shift and the Hubble law in  the Hoyle-Narlikar cosmology \cite{N}
\be
\label{73}
Z(D_c)=\frac{{\vh}(T)}{{\vh}(T-D_c/c)}-1\simeq
{\cal H}^c_{Hub}D_c/c;~~~~~~~~~~\,{\cal H}^c_{Hub}
=\frac{1}{{\vh}(T)}\frac{d\vh(T)}{dT}=
\frac{\sqrt{\rho(T)}}{\vh(T)}
\ee
reflect the change of the size of atoms in the process of evolution
of masses ~\cite{N,kpp,grg,plb}.

Equation (\ref{73})
gives the relation
between the present-day value of the scalar field and
cosmological observations  (the density of matter and the Hubble parameter)
\be
\label{d12a}
\vh(T)=\frac{\sqrt{\rho(T)}}{{\cal H}^c_{Hub}(T)}.
\ee
Note that the present-day observational data~\cite{rpp}
on the matter density
\be\label{d30}
\rho=\rho_b=\Omega_0\rho_{cr};~~~~~~~~~~~~~
0.1<\Omega_0<2~~~~~~~~~~~
(\rho_{cr}=\frac{3{\cal H}^c_{Hub}}{{8\pi}}M_{Pl}^2  )
\ee
give the value of the dynamic evolution parameter which
coincides with the Newton constant (or the Planck mass)
\be \label{exp}
\vh_0(T_0)=\mu\Omega_0^{1/2}.
\ee
Both the standards of measurement of the present day value
of $\vh_0$ in observational cosmology give the value of the Planck
mass~(\ref{exp}).
Nevertheless, only for the relative standard of a Weyl observer,
local measurements of the invariant interval do not depend on
the parameters of global evolution of the universe.

\section{"Measurable" Quantum Universe}

We calculate the generating functional for the unitary
perturbation theory as the S-matrix element in the standard
interaction representation applied in  quantum field theory
$$%\be \label{tr1}
S[\vh_1,\vh_2]=
\alpha^+<{\mbox out}~(\vh_2)|
T \exp\left\{-i\int\limits_{\vh_1}^{\vh_2}d\vh (H^R_I)\right\}
|(\vh_1)~{\mbox in}>+
$$
$$
\alpha^-<{\mbox out}~(\vh_1)|
\widetilde{T} \exp\left\{+i\int\limits_{\vh_1}^{\vh_2}d\vh (H^R_I)\right\}
|(\vh_2)~{\mbox in}>,
$$%\ee
where $T,~\widetilde{T}$ are symbols of ordering and anti-ordering,
$H^R_I$ is the Hamiltonian of interaction of the reduced system
\be \label{int}
H^R_I = H^R-H^R_0
\ee
$H^R$ is the reduced Hamiltonian defined by equation~(\ref{gde}),
$H^R_0$ is a free part of this Hamiltonian $H^R$
in the perturbation theory~(\ref{pt}),~(\ref{pert}),
and $|(\vh)~ {\mbox in (out)}>$
is the  ${\mbox in~ (out)}$ -
state of  Quantum Universe
which satisfies the Schr\"odinger equation
\be \label{sch}
\frac{d}{id\vh}|(\vh)~ {\mbox in (out)}>=H^R_0|(\vh)~ {\mbox in (out)}>.
\ee
As we have seen above, the  description of both the Dirac dynamics
(in the reduced phase space) and the measurable geometry, i.e.
the invariant interval
\be \label{perint}
(ds_c)^2=dT^2-(\delta_{ij}+ h_{ij})dx^idx^j
\ee
can be given only by
 the constrained system with the extended action
\be \label{exho}
W^E_0=\int\limits_{t_1}^{t_2} dt \left\{\left[\int d^3x \pi^T_{ij}
\dot h^T_{ij}
\right] -P_0\dot \vh - N_0\left[ -\frac{P_0^2}{V_0}+\bar H_0\right]\right\} ,
\ee
where
\be \label{ho}
\bar H_0=\int d^3x\left(\frac{6(\pi^T_{(h)})^2}{\vh^2}+\frac{\vh^2}{24}
(\ik_ih^T)^2\right);~~(h^T_{ii}=0;~~\ik_jh^T_{ji}=0),
\ee
is the "measurable" Hamiltonian of "free" gravitons.
This action includes  the world time interval $dT=N_0dt$
measured by an Weyl observer.

The dependence of the world time on the internal evolution parameter $\vh$
is treated as evolution of a classical Universe.
Quantization of the extended constrained system with the "free"
gravitons ~(\ref{exho}) was performed in paper ~\cite{ps1} where
the holomorphic variables of "particles" ($a^+,~a$)
were defined as  variables which
diagonalize the measurable Hamiltonian ~(\ref{ho}), and
"quasiparticles" ($b^+,~b$), as  variables
  which diagonalize the classical and quantum equations
of motion and lead to the equivalent oscillator-like
system with the set of conserved "quantum numbers".
For the latter system there is the canonical transformation
~\cite{lc,gkp,ps1}
of the
extended system ~(\ref{exho}) to a new set of variables
\be \label{levi}
(a^+,a| P_0,\vh) \Rightarrow (b^+,b| \Pi, \eta),
\ee
so that the new internal evolution parameter $\eta$ coincides,
in the equation of motion for the new momentum  $\Pi$,
with the invariant time measured by a Weyl observer in the comoving frame
of references
\be \label{eta}
\frac{\delta W_0^E}{\delta \Pi} = 0 \Rightarrow d\eta=N_0dt=dT.
\ee
Thus, after the canonical transformations
the states of the measurable Quantum Universe (with a conserved number of
"quasiparticles") are determined by the Schr\"odinger equation
\be \label{schp}
\frac{d}{id T}|\vh(T)~ {\mbox in}>
=\sum\limits_{n}\omega_{b}(n,T)
\frac{1}{2}(b_n^+b_n + b_nb_n^+)|\vh(T)~ {\mbox in}>
=E^R_0|\vh(T)~ {\mbox in}>,
\ee
where $n$ denotes a set of parameters of gravitons
(projections of spins, and momenta), and $E_0^R$ is an eigenvalue of the
oscillator-like Hamiltonian.

The state of "nothing" is
the squeezed vacuum (without quasiparticles)
$b_n(T)|\vh(T)~ {\mbox in}>=0$.
It was shown ~\cite{ps1} that for small $\vh$, and a large Hubble parameter,
at the beginning of the Universe, the state of  vacuum of quasiparticles
leads to the measurable density
\be \label{early}
{}_b<\rho(a^+,~a)>_b = \rho_0 \frac{1}{2} \left(\frac{\vh_0^2}{\vh^2(T)} +
\frac{\vh^2(T)}{\vh_0^2}\right),
\ee
where $\vh_0$ is the initial value, and
$\rho_0=\frac{1}{2}\sum\limits_{n}\omega_a(n)$
is the density of the Kasimir vacuum of "particles".
The first term corresponds to the rigid state equation
(in accordance with the
classification of the standard cosmology) and it
leads to the Kasner anisotropic stage $T_{\pm}(\vh)\sim \pm\vh^2$
(described by the Misner wave function ~\cite{M}).
>From the point of view of fields of matter for which $\vh$ forms masses,
the negative solution $ \vh^2(T_-)< 0$ (anti-Universe) is not stable, in
this stage.
The second term of the squeezed vacuum density~(\ref{early}) leads to
the stage with the inflation of the scale $\vh$ with respect to the time
measured by a Weyl observer
$$
\vh(T)\simeq \exp(T\sqrt{2\rho_0}/\vh_0).
$$
It is the
stage of intensive creation of "measurable particles".
After the inflation,  the Hubble parameter goes to zero, and
 gravitons convert into photon-like oscillator excitations with
the conserved number of particles.

At the present-day stage, we can describe
${\mbox in}$- and ${\mbox out}$ -states in terms of
the "measurable" time $T$ and the Hamiltonian  $\bar H_0$ ~(\ref{ho})
where $\vh$ is changed by $\mu$,
in agreement with the data of the observational cosmology
$\vh(T_0)=\mu$ discussed above.

The internal evolution parameter can be connected with the time
measured by an observer of a quantum state of the Universe
$|out>$ in terms
of the conserved quantum numbers of this state:
energy $E_{out}$ and density $\rho_{out}=E_{out}/V_0$
\be \label{tvh}
\frac{d\vh}{dT}=\sqrt{\rho_{out}}.
\ee
It is natural to suppose that $E_{out}$ is a tremendous energy in
comparison  with possible  deviations of the free Hamiltonian in
the laboratory  processes
\be \label{lab1}
 \bar H_0 = E_{out} + \delta H_0 ,~~~
<{\mbox out}| \delta H_0|{\mbox in}> << E_{out}.
\ee

\section{Infinite volume limit of  Quantum Gravity}

We consider the infinite volume limit of
the S-matrix element  in terms
of the measurable time $T$ for the present-day
stage  $ T=T_0$
taking into account only the contribution of the Universe
$\alpha^+=1, \alpha^-=0$
\be \label{tr2}
S[T_1=T_0-\Delta T|T_2=T_0+\Delta T]=<{\mbox out}~(T_2)|
T \exp\left\{-i\int\limits_{\vh(T_1)}^{\vh(T_2)}d\vh
(H^R_I)\right\}|(T_1)~{\mbox in}>.
\ee
One can express this matrix element in terms of the time measured by
an observer of an out-state with the tremendous number of particles
in the Universe using equation ~(\ref{tvh}) and the
approximation ~(\ref{lab1}).

In the infinite volume limit, we get
\be \label{limit}
d \vh [ H^R_I ]=dT[\hat F \bar H_I + O(1/V_0)+ O(1/E_{out})]
\ee
where $\bar H_I$ is the Hamiltonian of interaction in GR,
and
\be \label{form}
\hat F = \sqrt{\frac{E_{out}}{E_{out} + \delta \bar H_0}}
\ee
is the multiplier which plays the role of a form factor for
physical processes  observed at  the "laboratory"  conditions
when the cosmic energy $E_{out}$ is much greater than the
deviation of the free energy
\be \label{lab}
\delta \bar H_0 = \bar H_0-E_{out};~~
\ee
due to creation and annihilation of
real and virtual particles in the laboratory experiments.
The measurable time of the laboratory experiments $T_2-T_1$ is much
smaller than the age of the Universe $T_0$, but it is much greater
than the reverse "laboratory"  energy  $ \delta$, so that the
limit
$$
\int\limits_{T_1 }^{T_2 } \Rightarrow \int\limits_{-\infty }^{+\infty }
$$
is valid.
We can get the conventional quantum field theory representation of
matrix element ~(\ref{tr2})
\be \label{tr3}
S[-\infty|+\infty]=<{\mbox out}|
T \exp\left\{-i\int\limits_{-\infty}^{+\infty} dT
\bar H_I\right\}
|{\mbox in}>,
\ee
if we neglect  the form factor ~(\ref{form}) which removes a set of
ultraviolet divergences.
This matrix element corresponds to the FP functional integral
\be \label{qft}
Z_{QFT}~ = ~\int D(\bar q,\bar \pi,\bar N^k)[FP]_s
\exp\left\{ i\bar W^E[\bar q|\mu]+\mbox{sources}\right\}.
\ee
where $\bar W^E[\bar q|\mu]$ is the initial action~(\ref{gr}) in terms of
the conformal-invariant time $T$ for $\bar N = 1$~(\ref{time}).

The main difference of the obtained generating functional
from the Faddeev-Popov-DeWitt one ~\cite{fp2,dw} is the absence
of the fourth gauge which fixes the determinant of the space metric
~\cite{fp2} or its momentum ~\cite{d1}.
In both the cases, these gauges contradict the
motion equations for these variables, as we have seen above,
in the context of the Dirac perturbation theory~\cite{d1}.

The result~(\ref{qft}) could be predicted from the very beginning, the
problem was to show the range of validity of the conventional
quantum field perturbation theory ~\cite{fp1,fp2} and its possibilities
for solution of problems of the Early Universe.

The relativistic covariance of the considered  scheme of quantization
can be proved in the infinite space-time on the level of
algebra of commutation relations of the generators of the
Poincare symmetry in perturbation theory by analogy with QED~\cite{zumino}.

>From the point of view of the quantum field theory limit,
the conformal variables and measurable quantities, including
the conformal time, are favorable,
and the Einstein General Relativity looks like a scalar version
of the Weyl conformal invariant theory, where the Weyl scalar field
forms both the Planck mass (in agreement with the present-day astrophysical
data) and masses of elementary particles~\cite{plb} (in agreement with the
principle of equivalence).

In the Weyl theory, the Higgs mechanism of the formation of particle masses
becomes superfluous and, moreover, it contradicts  the equivalence
principle, as, in this case, the Planck mass and masses of
particles are formed by different scalar fields.

In the conformal theory~\cite{grg,plb},
we got the $\sigma$-version of the Standard Model~\cite{pr}
without  Higgs particles, and with the prescription~(\ref{form})
to be free from the ultra-violet divergences for the precision
calculations.

\section{ Conclusion}

We have obtained the generalization of the unitary S-matrix in General
Relativity ~\cite{dw,fp2,kp} for a finite space-time in agreement with
the group of invariance of the Hamiltonian dynamics in GR.
This group contains reparametrizations of the coordinate time $(t)$
and gauge transformations with three local parameters.
We have shown that the solution of one global constraint (with respect
to the zero Fourier harmonic of the space metric determinant $\vh_0$)
and three local constraints remove all ambiguities from the perturbation
theory for transverse gravitons, so that the fourth  gauge ~\cite{d1,fp2} for
fixation of the space metric determinant is superfluous and can contradict
equations of motion.

As a result of the solution of these constraints,
we got the unconstrained version of GR which describes
the dynamics
of the Dirac "observables" in the reduced phase space
with the dynamic evolution parameter. Besides the unconsrained dynamics,
the extended Hamiltonian GR contains the geometry of "measurable
quantities" (which depend on all components of metric
including those which cannot be defined by complete set of equations
in the sector of the Dirac "observables").

The geometric sector of "measurable intervals" is a specific feature of
GR which strongly distinguishes it from  classical unconstrained
systems where the dynamic evolution parameter coincides with
the measurable time.

In particular, the evolution of the universe is the
evolution of the  Dirac sector of "observables" (together with their
dynamic evolution parameter) with respect to the "measurable"
interval (including the invariant proper time),
and this "measurable" evolution goes beyond the scope of the
sector of the Dirac "observables". This fact is the main difficulty for
the standard quantization.

To emphasize the autonomy of the "measurable" geometrical sector in GR, we
pointed out two different standards of measurement
(relative and absolute) which correspond to
two theories with the same dynamics: GR and the scalar version of the
Weyl geometry of similarity (with
a scalar field as the measure of a change of the
length of a vector in its parallel transport).
In terms of the conformal invariant variables, actions of both
these theories coincide, but the measurable intervals are different.
An Einstein observer (who measures lengths by the absolute
standard) sees the Friedmann-Robertson-Walker evolution of a universe,
while a Weyl observer
(who treats the determinant of the three-dimensional metric multiplied by
the Planck constant as a measure of a change of the length of a vector
in its parallel transport) sees the Hoyle-Narlikar evolution.

We have considered the
phenomenon of Quantum Universe mainly "with a view to its measurement
describing the methods of measurement and defining the standards on which
they depend" ~\cite{mx}.

The phenomenon of Quantum Universe can be described by two measurable
quantities: the time in the comoving frame, and the red shift of spectral
lines of the cosmic object atoms
in terms of the dynamic evolution parameter (i.e. the scale factor).
Both these quantities determine the background metric of the
considered perturbation theory for the unconstrained GR
and  measurable density.

Now, we can define the Quantum Universe as the universe
filled by "free" quantum fields
in the space-time with the considered background metric and standard of
the measurement of the invariant time intervals.
The evolution of the Quantum Universe is expressed in terms of the
measurable time by  canonical transformations which convert
the dynamic evolution parameter into the measurable time and
the variables of particles (diagonalizing the measurable density) into the
quasiparticles (diagonalizing equations of motion) with the squeezed vacuum.

The Quantum Gravity is the theory of S-matrix between the states of the
Quantum Universe.

The infinite space-time limit of this S-matrix leads to the
standard quantum field theory S-matrix provided the measurable time is the
conformal time of a Weyl observer and General Relativity is the
scalar version of the Weyl conformal invariant theory with the set of
prediction, including

the Hoyle-Narlikar version of observational
cosmology, where the physical reason of red-shift is changing masses of
elementary particles in the process of evolution of the Universe,

the cosmic mechanism of the formation of both the masses of elementary
particles and the Planck mass by the Weyl scalar field
(which does not contradict the present-day astrophysical data),

the squeezed vacuum inflation from "nothing" at the beginning of the Universe,

and the negative
result of CERN experiment on the search of Higgs particles.

{\bf Acknowledgments}

\medskip

We are happy to acknowledge interesting and critical
discussions with  Profs.
A. Borowiec, A.V. Efremov,
G.A. Gogilidze, V.G. Kadyshevsky,
A.M. Khvedelidze, E.A. Kuraev,
D. Mladenov,
Yu.G. Palii, V.V. Papoyan,  and G.M. Vereshkov.
One of the authors (V.P.) thank Profs. A. Ashtekar,
C. Isham, T. Kibble, J. Lukierski,
L. Lusanna,  C. Rovelli for useful discussions. The work was supported
by the Committee for Scientific Researches grant no. 603/P03/96 and by
the Infeld-Bogolubov program.

\end{document}